\documentclass[%
 reprint,
 amsmath,amssymb,
 aps,
 showpacs,
pra,
]{revtex4-1}

\usepackage{graphicx}
\usepackage{dcolumn}
\usepackage{bm}

\newcommand{\be}{\begin{equation}}
\newcommand{\ee}{\end{equation}}
\newcommand{\bea}{\begin{eqnarray}}
\newcommand{\eea}{\end{eqnarray}}

\begin{document}
\setlength{\unitlength}{1.2mm}
\title{Renormalization group analysis of $p$-orbital Bose-Einstein condensates in a square optical lattice}
\author{Boyang Liu}
\affiliation{Beijing National Laboratory for Condensed Matter
Physics, Institute of Physics, Chinese Academy of Sciences, Beijing
100190, China}
\author{Xiao-Lu Yu}
\affiliation{Beijing National Laboratory for Condensed Matter
Physics, Institute of Physics, Chinese Academy of Sciences, Beijing
100190, China}
\author{Wu-Ming Liu}
\affiliation{Beijing National Laboratory for Condensed Matter
Physics, Institute of Physics, Chinese Academy of Sciences, Beijing
100190, China}

\date{\today}
\begin{abstract}
We investigate the quantum fluctuation effects in the vicinity of
the critical point of a $p$-orbital bosonic system in a square
optical lattice using Wilsonian renormalization group, where the
$p$-orbital bosons condense at nonzero momenta and display rich
phases including both time-reversal symmetry invariant and broken
BEC states. The one-loop renormalization group analysis generates
corrections to the mean-field phase boundaries. We also show the
quantum fluctuations in the $p$-orbital system tend to induce the
ordered phase but not destroy it via the the Coleman-Weinberg
mechanism, which is qualitative different from the ordinary quantum
fluctuation corrections to the mean-field phase boundaries in
$s$-orbital systems.  Finally we discuss the observation of these
phenomena in the realistic experiment.
\end{abstract}
 \maketitle

\section{Introduction}
Confining cold atoms in an optical lattice
has proven to be an exciting and rich environment for studying many
areas of physics \cite{Jaksch,Greiner,Muller,Wirth,Panahi}. However,
the ground state wavefunctions of single component bosons are
positive defined in the absence of rotation as described in the
``no-node" theorem \cite{Feynman}, which imposes strong constraint
for the feasibility of using boson ground states to simulate
many-body physics of interest. One way of circumventing this
restriction is to consider the high orbital bands since the
``no-node'' theorem only applies to ground states \cite{Wu}. The
unconventional Bose-Einstein condensations (BECs) of high orbital
bosons exhibit more intriguing properties than the ordinary
BECs, 
including the nematic superfluidity \cite{Isacsson,Xu}, orbital
superfluidity with spontaneous time-reversal symmetry breaking
\cite{Liu,Wu1,Stojanovi,Kuklov,Cai,Lewenstein,martikainen} and other
exotic properties \cite{Bergman,Scarola,Li}. The theoretic work on
the $p$-orbital fermions is also exciting
\cite{Wu2,Kai,Zhao,Shizhong,Li2,Zhang}. Furthermore, the $p$-orbital
and multi-orbital superfluidity have been recently realized
experimentally by pumping atoms into high orbital bands
\cite{Strabley,Muller,Wirth,Olschlager,Panahi}.

Since the $p$-orbital Bose gas exhibits rich phase structures, it is
interesting to investigate the quantum fluctuation effects in the
vicinity of the critical point in presence of competing orders. Of
particular interest is the quantum fluctuation induced symmetry
breaking (QFISB). This phenomenon was first discussed by S. Coleman
and E. Weinberg \cite{Coleman,Amit}. They investigated a theory of a
massless charged meson coupled to the electrodynamic field by
effective potential method. Starting from a model without symmetry
breaking at tree level they found that at one-loop level a new
energy minimum was developed away from the origin, thus, the $U(1)$
symmetry of the complex scalar field is spontaneously broken.
Independently, Halperin, Lubensky, and Ma \cite{Halperin} discovered
the analogous phenomenon in the Ginzburg-Landau theory of
superconductor to normal metal transition. Furthermore, this quantum
fluctuation induced phase transition is found to be first-order
\cite{Amit}. Recently, there have appeared more examples that the
symmetry of certain order parameters can be spontaneously broken by
the quantum fluctuations in condensed matter systems
\cite{Continentino,Ferreira,Qi,Millis,She,Diehl,Bonnes}. For
instance, in the system of lattice bosons with a three-body
hard-core constraint the transitions between the dimer superfluid
phase and the conventional atomic superfluid state are proposed to
be Ising-like at unit filling and driven first-order by fluctuations
via the Coleman-Weinberg mechanism at other fractional filling
\cite{Diehl,Bonnes}.

In this paper, we study the  $p$-orbital bosonic system in a square
optical lattice using renormalization group (RG) analysis. The
spectrum of $p$-orbital bosons in square lattice has two energy
minima in the Brillouin zone located at $\vec K_X=(\frac{\pi}{a},0)$
for $p_x$-band and $\vec K_Y=(0,\frac{\pi}{a})$ for $p_y$-band
respectively \cite{Wirth,Wu}. A macroscopic number of the
$p$-orbital bosons can condense at these two energy minima. This
phenomenon is usually named as ``unconventional BEC" \cite{Wu,Cai}.
At these two band minima the Bloch wave functions are time-reversal
invariant and, thus, real valued. Lattice asymmetry favors a ground
state that bosons condense at either $\vec K_X$ or $\vec K_Y$, which
is called real BECs. A linear superposition of these two real valued
wave functions with a fixed phase difference forms a complex BEC,
which is favored by the system with inter-species interactions
between $p_x$ and $p_y$ orbital bosons and spontaneously breaks the
time-reversal symmetry \cite{Cai, Wu}. Since the unconventional BEC
is beyond the constraint of ``no-node" theorem, it has intriguing
properties. In our work we use RG to investigate the quantum phase
transitions between the real and complex BEC phases. We find that
when the inter-species interactions are turned on the real BEC
phases may become unstable and the system can finally flow to the
complex BEC phase. The phase transitions for the real BEC phases to
the complex BEC phase require the $U(1)$ symmetry breaking of $p_x$
or $p_y$ orbital bosons. This is a phenomenon of quantum fluctuation
induced phase transition. The phase transitions induced the quantum
fluctuations have proven to be in first order \cite{Amit}, which has
different scaling behaviors from the second-order ones
\cite{Nienhuis,Fisher1}. Based on the recent researches on the
quantum criticality in coldatom physics
\cite{Zhou,Hazzard,Zhang1,Donner,Zhang2}, we can propose a method to
observe this first-order phase transition in the realistic
experiment.

\section{The renormalization group flow equations and phase
diagrams}

\begin{figure}[h]
  \includegraphics[width=7.5cm]{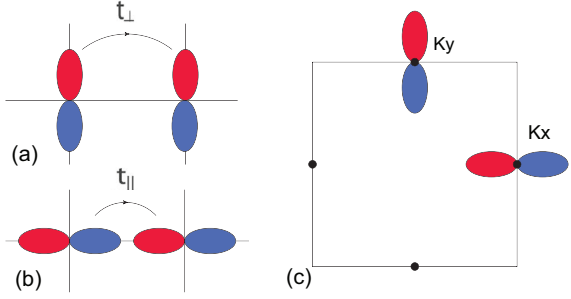}
  \caption{(Color online) The bonding pattern of p-orbitals: (a) the
$\sigma$-bonding and (b) the $\pi$-bonding. (c) The p-orbital band structure in a square lattice. The band minima are located at $\vec K_X=(\frac{\pi}{a},0)$ for the $p_x$-orbital band, and
$\vec
K_Y=(0,\frac{\pi}{a})$ for the $p_y$-orbital band, respectively. } \label{fig:lattice}
 \end{figure}

The tight-binding model of the $p$-orbital bosons in a square
lattice is described by a Hamiltonian as following \bea
&&H=t_\|\sum_{<ij>}[p^\dagger_{i,\hat e_{ij}}p_{j,\hat
e_{ij}}+h.c.]-t_\bot\sum_{<ij>}[p^\dagger_{i,\hat f_{ij}}p_{j,\hat
f_{ij}}+h.c.],\cr&&\eea where $\hat e_{ij}$ and $\hat f_{ij}$ are
two unit vectors. $\hat e_{ij}$ is along the bond orientation
between two neighboring sites $i$ and $j$ and $\hat f_{ij}=\hat
z\times \hat e_{ij}$. $p_{\hat e_{ij}}$ and $p_{\hat f_{ij}}$ are
the projections of p-orbitals along and perpendicular to the bond
direction respectively. The $\sigma$-bonding $t_\|$ and the $\pi$
bonding $t_\perp$ describe the hoppings along and perpendicular to
the bond direction, which are illustrated in graph (a) and (b) of
Fig.\ref{fig:lattice}.
This tight-binding model shows that the energy minima are located at
half values of the reciprocal lattice vectors \cite{Wu, Wirth} as
depicted in Fig.\ref{fig:lattice}(c), where $\vec
K_X=(\frac{\pi}{a},0)$ and $\vec K_Y=(0,\frac{\pi}{a})$. The
$p$-orbital bosons can condense at either $K_X$ or $K_Y$ to form a
real BEC or both of $K_X$ and $K_Y$ to form a complex BEC. To
describe the phase transitions we write down a Landau-Ginzberg
theory. The action with the most general interactions is cast as
following \bea \mathcal {Z}=&&\int
D[\phi^\ast_{1},\phi_{1},\phi^\ast_{2},\phi_{2}]e^{-S[\phi^\ast_{1},\phi_{1},\phi^\ast_{2},\phi_{2}]},\eea
where \bea && S[\phi^\ast_1,\phi_1,\phi^\ast_2,\phi_2]=\cr
&&\int^\infty_{-\infty} \frac{d\omega}{2\pi}\int^\Lambda_0 \frac{d^2
k}{(2\pi)^2} \sum_{i=1,2}\phi_i^\ast(\omega, \vec
k)(-i\omega+\epsilon_k- r_i)\phi_i(\omega, \vec k)\cr &&
+\int_{\omega k}^\Lambda\Big\{
\sum_{i=1,2}g_i\phi_i^\ast(\omega_4,\vec
k_4)\phi_i^\ast(\omega_3,\vec k_3)\phi_i(\omega_2,\vec
k_2)\phi_i(\omega_1,\vec k_1)\cr&&+g_3\phi_2^\ast(\omega_4,\vec
k_4)\phi_1^\ast(\omega_3,\vec k_3)\phi_1(\omega_2,\vec
k_2)\phi_2(\omega_1,\vec k_1)\cr&&+g_4[\phi_1^\ast(\omega_4,\vec
k_4)\phi_1^\ast(\omega_3,\vec k_3)\phi_2(\omega_2,\vec
k_2)\phi_2(\omega_1,\vec k_1)+H.C.]\Big\},\cr&&\eea where $\phi_1$
and $\phi_2$ describe the condensate order parameters at $K_X$ and
$K_Y$, respectively. Here we used a short-handed notation
$\int_{\omega k}^\Lambda=\prod^4_{i=1}\int^\infty_{-\infty}
\frac{d\omega_i}{2\pi}\int^\Lambda_0\frac{d^2
k_i}{(2\pi)^2}(2\pi)^2\delta(\vec k_4+\vec k_3-\vec k_2-\vec
k_1)\cdot(2\pi)\delta(\omega_4+\omega_3-\omega_2-\omega_1)$. A
cutoff $\Lambda$ is given to the momentum space since this is a low
energy effective theory.  Terms with $g_1$ and $g_2$ are the
intra-species interactions of $p_x$ and $p_y$ orbital bosons. Terms
with $g_3$ and $g_4$ describes the inter-species interactions. The
$g_4$ term can rise in this high orbital model as an Umklapp
scattering process since the momentum transfer is $\pm2(\vec
K_X-\vec K_Y)$, which equals to the reciprocal lattice vectors
\cite{Cai, Wu}.

In contrast to the mean-field theory where certain order parameter
is presumably defined, the renormalization group analysis treats
various instabilities on an equal footing without assuming any
specific order parameters. Here we implement momentum-shell
renormalization group method to study the running of various
parameters. Following the Wilson's approach \cite{Wilson} the
renormalization group transformation involves three steps: (i)
integrating out all momenta between $\Lambda/s$ and $\Lambda$, for
tree level analysis just discarding the part of the action in this
momentum-shell; (ii) rescaling frequencies and the momenta as
$(\omega, k)\rightarrow (s^{[\omega]}\omega, sk )$ so that the
cutoff in k is once again at $\pm\Lambda$; and finally (iii)
rescaling fields $\phi \rightarrow s^{[\phi]} \phi$ to keep the
free-field action $S_0$ invariant.

\begin{figure}[h]
  \includegraphics[width=6.5cm]{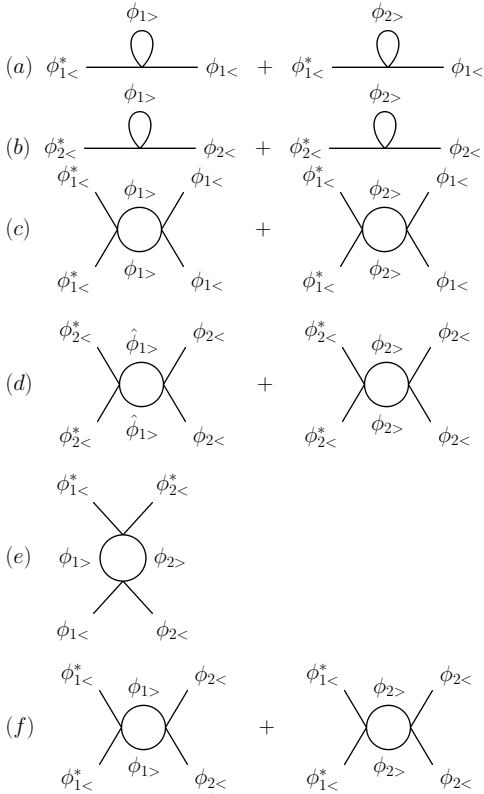}
  \caption{ The one-loop Feynman graphs contributing to the renormalization of
(a) the parameter $r_1$, (b) the parameter $r_2$, (c) the interaction $g_1$, (d) the interaction $g_2$, (e) the interaction $g_3$ and (f) the interaction $g_4$.} \label{fig:feynman}
 \end{figure}

In order to perform the first step of Wilson's approach in one-loop
level, we need to split the fields into ``slow modes"
$\phi_{i<}(\omega, \vec k)$ and ``fast modes" $\phi_{i>}(\omega,
\vec k)$. Then we have \be \phi_i(\omega, \vec k)=\phi_{i<}(\omega,
\vec k)+\phi_{i>}(\omega, \vec k),\ee where
 \bea
&&\phi_{i<}(\omega, \vec k)~~~~\mbox{for}~~~~0<|k|<\Lambda/s,\cr&&
\phi_{i>}(\omega, \vec
k)~~~~\mbox{for}~~~~\Lambda/s<|k|<\Lambda.\eea The partition
function now can be written as \bea &&\mathcal {Z}=\cr &&\int
D[\phi^\ast_{1<},\phi_{1<},\phi^\ast_{2<},\phi_{2<}]e^{-S[\phi^\ast_{1<},\phi_{1<},\phi^\ast_{2<},\phi_{2<}]}\cr
&&\times\int D[\phi^\ast_{1>},\phi_{1>},\phi^\ast_{2>},\phi_{2>}]\cr
&&
e^{-S_0[\phi^\ast_{1>},\phi_{1>},\phi^\ast_{2>},\phi_{2>}]-S_I[\phi^\ast_{1<},\phi_{1<},\phi^\ast_{2<},\phi_{2<},\phi^\ast_{1>},\phi_{1>},\phi^\ast_{2>},\phi_{2>}]}.\cr&&\eea
We next construct an effective action by integration over the fast
fields. To the one-loop order, one obtains\bea
&&e^{-S_{eff}[\phi^\ast_{1<},\phi_{1<},\phi^\ast_{2<},\phi_{2<}]}\cr
&&=e^{-S[\phi^\ast_{1<},\phi_{1<},\phi^\ast_{2<},\phi_{2<}]}\cr
&&\cdot
\exp\Big[-\big<S_I[\phi^\ast_{1<},\phi_{1<},\phi^\ast_{2<},\phi_{2<},\phi^\ast_{1>},\phi_{1>},\phi^\ast_{2>},\phi_{2>}]\big>_>\cr
&&+\frac{1}{2}\big<S_I[\phi^\ast_{1<},\phi_{1<},\phi^\ast_{2<},\phi_{2<},\phi^\ast_{1>},\phi_{1>},\phi^\ast_{2>},\phi_{2>}]^2\big>_>\Big],\eea
where $\big<...\big>_>$ denotes the average over the fast
fluctuations. we perform the integrals over the fast modes by
evaluating the appropriate Feynman diagrams contributing to the
renormalization of the vertices of interest. The one-loop Feynman
graphs contributing to the renormalization are shown in Fig.
\ref{fig:feynman}.
One finds that the parameters $r_1$ and $r_2$ scale according to the
following relations up to one-loop order: \bea && \frac{d\tilde
r_1}{d\ell}=2\tilde r_1+4\tilde g_1 \cdot\theta(\tilde
r_1-1/2)+\tilde g_3\cdot \theta(\tilde r_2-1/2),\cr && \frac{d\tilde
r_2}{d\ell}=2\tilde r_2+4\tilde g_2 \cdot\theta(\tilde
r_2-1/2)+\tilde g_3\cdot \theta(\tilde r_1-1/2).\label{mu}\eea In
above equations we defined the dimensionless parameters as $\tilde
r_i= r_i m/\Lambda^2$ and $\tilde g_i=g_i m/(2\pi)$. The $\theta$
functions in the flow equations of $\tilde r_1$ and $\tilde  r_2$
are from $\omega$ integrations in the one-loop calculations, where
we have $\theta(\tilde
r-1/2)=\int^{+\infty}_{-\infty}\frac{d\omega}{2\pi}\frac{e^{i\omega
0^+}}{i\omega-(1/2-\tilde r)}$. Notice that we introduced the factor
$e^{i\omega 0^+}$ into the $\omega$ integral otherwise the integral
over $\omega$ doesn't converge \cite{Shankar}. Most of the studies
investigated the behaviors around the critical point, then these
contributions can be ignored at zero temperature
\cite{Sachdev,Dunkel}. However, in our case the running behaviors of
$ r_1$ and $ r_2$ to $+\infty$ or $-\infty$ are used as the criteria
of which phase the system will fall into. Therefore, we have to take
these contributions into account.  They will give severe influence
to the running of $\tilde r_1$ and $\tilde r_2$. The flow equations
of the coupling constants are as following: \bea && \frac{d\tilde
g_1}{d\ell}=-2\tilde g^2_1-2\tilde g_4^2 ,\cr && \frac{d\tilde
g_2}{d\ell}=-2\tilde g^2_2-2\tilde g_4^2 ,\cr && \frac{d\tilde
g_3}{d\ell}=-\tilde g_3^2 ,\cr &&  \frac{d\tilde
g_4}{d\ell}=-2\tilde g_1\tilde g_4 -2\tilde g_2\tilde g_4.\eea   All
the coupling constants with positive initial values are marginally
irrelevant. For example, $\tilde g_3$ can be solved as $\tilde
g_3(\ell)=\frac{\tilde g_3(0)}{1+\tilde g_3(0)\ell}$. It approaches
to zero as the length scale $\ell$ goes to infinity. However, it
doesn't imply that we can ignore these irrelevant coupling
constants. This is because the small $\tilde g_i$ will generate
small contributions to $\tilde  r_i$, which will then quickly grow
under the renormalization. As discussed by R. Shankar
\cite{Shankar}, an irrelevant operator can modify the flow of the
relevant couplings before it renormalizes to zero.

\begin{figure}[h]
  \includegraphics[width=7.5cm]{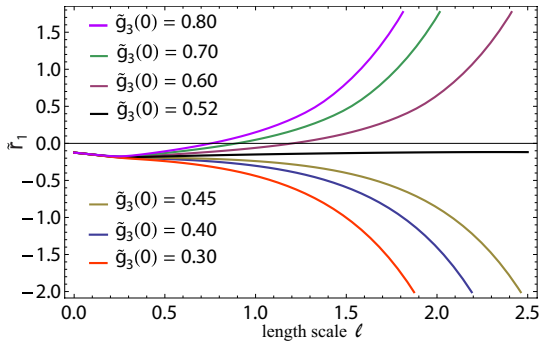}
  \caption{(Color online) The flow directions of parameter $\tilde r_1(\ell)$ with different interaction coupling $\tilde g_3(0)$.
  As $\tilde g_3(0)$ is increased the running direction of $\tilde r_1(\ell)$ changes from negative infinity to positive
  infinity. The system can finally flow to a condensed phase.
  The initial values of the parameters are $\tilde r_1(0)=-0.125$, $\tilde r_2(0)=0.3$
  and $\tilde g_1(0)=\tilde g_2(0)=\tilde g_4(0)=0.1$.  } \label{fig:running}
 \end{figure}

The running parameters $\tilde  r_1(\ell)$ and $\tilde  r_2(\ell)$
are relevant and can be solved numerically from Eq.(\ref{mu}). We
find that in regions of $\tilde  r_1(0)\geq\frac{1}{2}~\&~\tilde
 r_2(0)\geq\frac{1}{2}$ the solutions are \bea &&\tilde
r_1(\ell)=e^{2\ell}\big[\tilde r_1(0)+\int^\ell_0 e^{-2t}(4\tilde
g_1(t)+\tilde g_3(t))dt\big], \cr &&\tilde
r_2(\ell)=e^{2\ell}\big[\tilde r_2(0)+\int^\ell_0 e^{-2t}(4\tilde
g_2(t)+\tilde g_3(t))dt\big].\eea In this region the initial values
$\tilde r_1(0)$ and $\tilde r_2(0)$ and the one-loop level
contributions are all positive since all the interactions are
repulsive. Thus, $\tilde r_1(\ell)$ and $\tilde r_2(\ell)$ are both
running to positive infinity fast due to the exponential prefactor.
The one-loop contributions don't change the flow directions of the
chemical potentials. The running directions of $\tilde  r_1(\ell)$
and $\tilde  r_2(\ell)$ are completely determined by their initial
values. If the initial values are positive or negative, they finally
flow to positive or negative infinity.  In region of $\tilde
r_1(0)\geq\frac{1}{2}~\&~\tilde r_2(0)<\frac{1}{2}$ the solutions
are \bea &&\tilde r_1(\ell)=e^{2\ell}\big[\tilde r_1(0)+\int^\ell_0
e^{-2t}4\tilde g_1(t)dt\big],\cr &&\tilde
r_2(\ell)=e^{2\ell}\big[\tilde r_2(0)+\int^\ell_0 e^{-2t}\tilde
g_3(t)dt\big].\eea In region of $\tilde r_1(0)<\frac{1}{2}~\&~\tilde
r_2(0)\geq\frac{1}{2}$ the solutions are \bea &&\tilde
r_1(\ell)=e^{2\ell}\big[\tilde r_1(0)+\int^\ell_0 e^{-2t}\tilde
g_3(t)dt\big],\cr &&\tilde r_2(\ell)=e^{2\ell}\big[\tilde
r_2(0)+\int^\ell_0 e^{-2t}4\tilde g_2(t)dt\big].\eea Different from the
first region, in these two regions $\tilde r_1(0)$ or $\tilde
r_2(0)$ can be negative. The positive contributions from the
one-loop graphs may qualitatively change the running behaviors of
$\tilde r_1(\ell)$ or $\tilde r_2(\ell)$. That is, even if $\tilde
r_1(\ell)$ or $\tilde r_2(\ell)$ runs to negative infinity at the
tree level, the positive one-loop contributions can make $\tilde
r_1(\ell)$ or $\tilde r_2(\ell)$ go to positive infinity eventually.
The system will finally end up in a different phase. This generate
critical lines these two regions, which are determined by conditions
$\tilde \mu_2(0)+\int^\infty_0 e^{-2t}\frac{\tilde g_3(0)}{1+\tilde
g_3(0)t}dt =0$ and $\tilde \mu_1(0)+\int^\infty_0
e^{-2t}\frac{\tilde g_3(0)}{1+\tilde g_3(0)t}dt =0.$ However, in
other regions the running directions of $\tilde  r_1(\ell)$ and
$\tilde r_2(\ell)$ can eventually be changed by the one-loop
corrections from the interaction couplings, even if they renormalize
to zero. For instance, in Fig. \ref{fig:running} we start the
running of $\tilde r_1(\ell)$ from a negative initial value.
As we
vary the interaction coupling $\tilde g_3(0)$ from $0.3$ to $0.8$ we
observe that the running of $\tilde r_1(\ell)$ can finally be
changed from the negative to the positive direction. That is, even
if $\tilde r_1(\ell)$ runs to negative infinity at the tree level,
the positive one-loop contributions can make $\tilde  r_1(\ell)$ go
to positive infinity eventually. The system will finally end up in a
different phase. In region of $\tilde r_1(0)<\frac{1}{2}~\&~\tilde
r_2(0)<\frac{1}{2}$ the solutions are \bea &&\tilde r_1(\ell)=\tilde
r_1(0)e^{2\ell},~~~~~~~~~~~\tilde r_2(\ell)=\tilde r_2(0)e^{2\ell}.\eea In
this region it's easy to see the running behaviors are totally
determined by the tree level scaling. However, with certain initial
values $\tilde r_1(\ell)$ and $\tilde r_2(\ell)$ can finally flow to
the second or the third region and then continuously flow to
positive infinity or negative infinity, there are also critical
lines in this region.

\begin{figure}[t]
  \includegraphics[width=8cm]{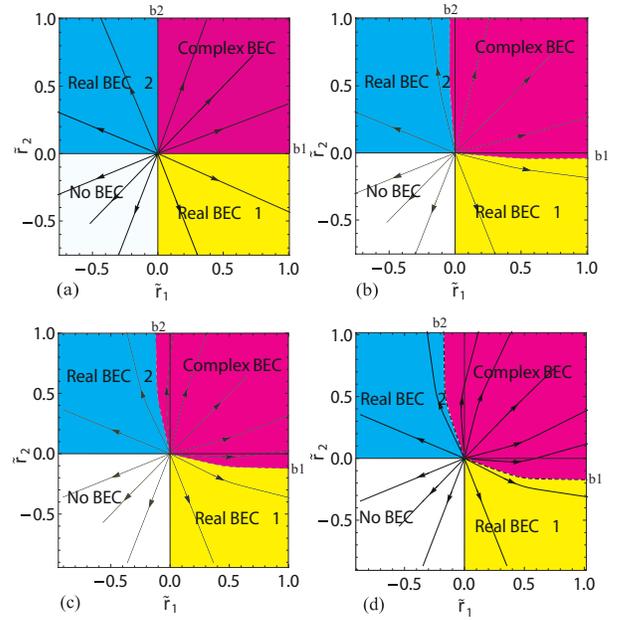}
  \caption{(Color online) Phase diagrams of the $p$-orbital
boson system. Four phases are determined by the flow directions of
the parameters $\tilde r_1$ and $\tilde r_2$ under the
renormalization group transformation. In the complex BEC phase
$\tilde  r_1(\ell)$ and $\tilde  r_2(\ell)$ run to positive
infinities, which represents that both $p_x$ and $p_y$ orbital
bosons condense. In the real BEC 1 (or 2) phase the $p_x$ (or $p_y$)
orbital boson condenses and the other one doesn't. In the no BEC
phase neither of the two orbital bosons condense. The initial values
of the couplings are $\tilde g_1(0)=\tilde g_2(0)=\tilde g_4(0)=0.1$
for all of the four diagrams. The inter-species interaction coupling
$\tilde g_3(0)$ between the $p_x$ and $p_y$ orbital bosons is set to
$0, 0.1, 0.3$ and $0.5$ for graphs (a), (b), (c) and (d)
respectively. ``b1" and ``b2" denote boundary 1 and boundary 2
between the real and complex BEC phases.} \label{fig:com}
 \end{figure}

Based on the numerical calculations the phase diagrams can be drawn
in Fig. \ref{fig:com}. The four phases are determined by the flow
directions of $\tilde r_1(\ell)$ and $\tilde  r_2(\ell)$ as
$\ell\rightarrow \infty$.
\begin{itemize}
\item[(I)] Complex BEC: $\tilde
 r_1(\ell)\rightarrow +\infty$ and $\tilde  r_2(\ell)\rightarrow
+\infty$,
\item[(II)] Real BEC 1 : $\tilde  r_1(\ell)\rightarrow
+\infty$ and $\tilde  r_2(\ell)\rightarrow -\infty$,
\item[(III)]
Real BEC 2: $\tilde  r_1(\ell)\rightarrow -\infty$ and $\tilde
 r_2(\ell)\rightarrow +\infty$,
\item[(IV)] No BEC: $\tilde
 r_1(\ell)\rightarrow -\infty$ and $ \tilde  r_2(\ell)\rightarrow
-\infty$.
\end{itemize}
We find that without inter-species interactions the complex BEC
phase is confined in the first quadrant of the phase diagram as
shown in (a) of Fig. \ref{fig:com}. However, as we turn on the
inter-species interactions the complex phase is enlarged into the
second and fourth quadrants as shown in (b), (c) and (d) of Fig.
\ref{fig:com}.
Comparing the four phase diagrams in Fig. \ref{fig:com}, we see that
as the inter-species interaction $\tilde g_3(0)$ becomes stronger
the complex BEC phase get enhanced. That is, certain region of real
BEC phases become unstable when the inter-species interactions are
turned on and the system finally flows to the complex BEC phase.
These interactions give rise to an instability from the real BEC
phase to the complex BEC phase. The phase transitions for the real
BEC phases to the complex BEC phase require the $U(1)$ symmetry
breaking of $p_x$ or $p_y$ orbital bosons. Given that $\tilde
g_3(0)$ is small, we can obtain the approximate expressions of the
two boundaries. Boundary 1 is \bea &&\tilde
 r_2(0)=-\frac{\tilde  r_1(0)\tilde
g_3(0)}{1-\frac{1}{2}\ln(2\tilde r_1(0))\cdot \tilde
g_3(0)}~\mbox{for}~0<\tilde r_1(0)<\frac{1}{2}\cr&& \tilde
r_2(0)=-\frac{\tilde g_3(0)}{2} ~\mbox{for}~\tilde
r_1(0)>\frac{1}{2}.\eea Boundary 2 is \bea&& \tilde
r_1(0)=-\frac{\tilde
 r_2(0)\tilde g_3(0)}{1-\frac{1}{2}\ln(2\tilde r_2(0))\cdot \tilde
g_3(0)}~\mbox{for}~0<\tilde r_2(0)<\frac{1}{2}\cr&&\tilde
r_1(0)=-\frac{\tilde g_3(0)}{2} ~\mbox{for}~\tilde
r_2(0)>\frac{1}{2}.\eea They are indicated by ``$b1$" and ``$b2$" in
Fig. \ref{fig:com}.

\section{Quantum fluctuation induced symmetry breaking}
\begin{figure}[h]
  \includegraphics[width=8cm]{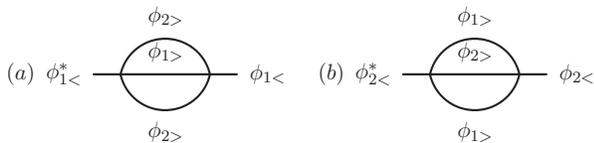}
  \caption{The Feynman diagrams of the two-loop corrections to the parameter (a) $ r_1$ and (b) $ r_2$. $\phi_1$ and $\phi_2$ are the boson fields defined in Eq. (1).}
  \label{fig:sunrise}
 \end{figure}

\begin{figure}[h]
  \includegraphics[width=6cm]{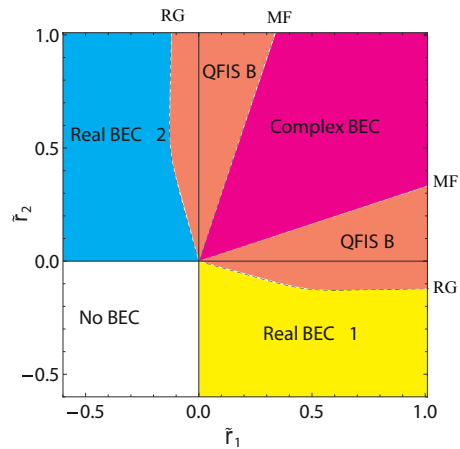}
  \caption{(Color online) Comparison of phase boundaries from the renormalization group analysis and the mean-field analysis. ``MF" and ``RG" indicate the boundaries from the mean-field and renormalization group
  analysis respectively. The complex BEC phase in
  renormalization group analysis is larger than the one from
  mean-field theory by a region named ``QFISB".
  The coupling constants are $\tilde g_1=\tilde g_2=0.5$,
  $\tilde g_3=0.3$ and $ \tilde g_4=0$.}
  \label{fig:qfisb}
 \end{figure}

The
mean-field results of the phase transition was derived by
constructing a Ginzburg-Landau theory in Ref. \cite{Cai}. However,
our renormalization group analysis gives some qualitative
differences: (I) In the mean-field analysis the boundary conditions
between the real and complex BEC phases depend on the
self-interaction couplings $g_1$ and $ g_2$. However, these two
couplings don't affect the phase boundaries in our results. The
quantum phase transition is purely induced by the inter-species
interaction between the $p_x$ and $p_y$ orbital bosons in RG
analysis. (II) At one-loop level the $g_4$ term doesn't give any
contributions to the flow equations of parameter $ r_1$ and $ r_2$.
It can get involved in higher order calculations. For instance,
$g_4$ term can generate corrections to the boundaries ``b1" and
``b2" at two-loop level through the sunrise graphs in Fig.
\ref{fig:sunrise}.
However, in mean-field analysis the phase boundaries depend on both
$g_4$ and $g_3$. This difference originates from the starting points
of the two analysis. The $g_4$ term explicitly breaks the original
$U(1)\times U(1)$ symmetry to $U_D(1)$ symmetry, where the index
``D" denotes for ``Diagonal", and leads to a fixed phase difference
between the two fields $\phi_1$ and $\phi_2$. The mean-field
analysis starts from this symmetry breaking phase. Hence, the
complex phase in mean-field is a coherent superposition of the two
ground states. However, our renormalization group analysis starts
from the normal phase and focuses on the effects of the quantum
fluctuations. In this case $g_4$ term doesn't show its contributions
up to one-loop level. Our complex phase is just a incoherent mixture
of the two ground states. (III) The comparison of the phase diagrams
of the renormalization group analysis and mean-field theory can be
illustrated in Fig. \ref{fig:qfisb}.
In order to compare the two phase diagrams with the same
circumstance we set $g_4=0$ in the mean-field phase diagram. The
expressions of the mean-field phase boundaries are $\frac{\tilde
r_2}{\tilde r_1}=\frac{2\tilde g_2}{\tilde g_3}$ and $\frac{\tilde
r_2}{\tilde r_1}=\frac{\tilde g_3}{2\tilde g_1}$ \cite{Cai}. In
Fig.\ref{fig:qfisb} it's obvious to see that the crucial difference
is that the complex BEC phase is enlarged and the real BEC phase is
suppressed in the RG phase diagram.

In essence, the above differences can be explained as effects of the
``quantum fluctuation induced symmetry breaking". The phase
transitions from real BEC to complex BEC phase indicate the $U(1)$
symmetry breakdown of field $\phi_1$ or $\phi_2$. Mean-field
description of the symmetry breaking is based on semiclassical
approximation. In other words, it's a tree-level result. When we
take into account the quantum fluctuation, the one-loop corrections
can significantly change the model parameters and make some region
of the real BEC phase become unstable. In Fig. \ref{fig:qfisb} these
regions are labeled by ``QFISB". The original QFISB effect was
studied using the effective potential method \cite{Coleman}. Our
work reaches a qualitatively analogous result using renormalization
group analysis.

Of particular importance is that this phenomenon has qualitative
difference from the ordinary quantum corrections to the mean-field
results. For instance, in a $s$-orbital system the phase transition
is described by $\phi^4$ theory. The renormalization group
calculations show that the system may flow from a ordered phase to a
disordered phase when the quantum fluctuations are taken into
account. That is, the quantum fluctuations tend to destroy the
ordered phase but not induce it \cite{Alexander}. However, in our
$p$-orbital system the ordered phase of one type of orbital bosons
may be induced by the quantum fluctuations from the interactions
with the other type of orbital bosons. In Fig. \ref{fig:com} we show
that the system in the disordered phase (real BEC phase) may flow to
the ordered phase (complex BEC phase) under the RG transformations.
Further more, in the $\phi^4$ theory of the $s$-orbital system the
phase transition occurs at the Wilson-Fisher fixed point with finite
values of couplings. The transition is second-order. However, in our
work $ r_1$ and $ r_2$ are both runaway trajectories. They flow to
$+\infty$ or $-\infty$. This infinity is characteristic of
first-order phase transition. Actually, the quantum fluctuation
induced phase transition has proven to be first-order in effective
potential method \cite{Amit}.

\section{Finite temperature scaling and experimental proposal} In
order to give a realistic experimental proposal to observe this
quantum fluctuation induced first-order phase transition we
investigate the finite temperature scaling behaviors of our system.
In the vicinity of the quantum critical points the observables obey
universal scaling relations. A interesting feature of the scaling
approach is that it allows to determine the singular behavior of the
physics quantities of interests as a function of temperature at
criticality. For instance, we consider the system undergoes a phase
transition between the ``complex BEC" phase and the ``real BEC 2"
phase. To discuss this transition, it's convenient to introduce a
parameter $\delta\equiv r_1-r_{1c}$ to measure the distance to the
transition. Then, the finite temperature scaling behavior of the
free energy density near the critical line ``b1" can be described as
following \bea f(\delta, T)\sim |\delta|^{\nu(d+z)}\tilde
f(T/|\delta|^{\nu z}),\eea where $d$ is the special dimension of the
system, $\nu$ is the correlation length exponent and $z$ is the
dynamic exponent. $\tilde f(u)$ is a universal scaling function,
which approaches a constant as $u\rightarrow 0$. Thus, the critical
temperature $T_c$ vanishes like $T_c=u_c \delta^{z\nu}$ for small
$\delta$. A general discussion shows that the scaling behaviors near
a first-order phase transition are characterized by the scaling
exponents such as $\beta=0$, $\alpha=\gamma=1$ and $\nu=1/(d + z)$
\cite{Nienhuis,Fisher1}. In our case, the dynamic exponent is $z=2$
and the system dimension is $d=2$. Thus, the correlation length
exponent is $\nu=1/(d + z)=1/4$. The scaling of the critical
temperature is $T_c\sim \sqrt{|\delta|}$.

Recently, several schemes were proposed to determine the critical
properties in cold-atom systems by extracting the universal scaling
functions from the atomic density profiles
\cite{Zhou,Hazzard,Zhang1}. The experimental observations of quantum
critical behavior of ultracold atoms have also been reported
\cite{Donner,Zhang2}. The study of quantum criticality in cold-atom
systems is based on \emph{in situ} density measurements
\cite{Zhou,Hazzard,Zhang1,Zhang2}. The density can be cast as
$n(\mu, T)-n_r(\mu,
T)=T^{\frac{d}{z}+1-\frac{1}{z\nu}}G(\frac{\mu-\mu_c}{T^{1/z\nu}})$,
where $\mu_c$ is the critical value of the chemical potential, $n_r$
is the regular part of the density and $G(x)$ is a universal
function describing the singular part of the density. Following the
scheme developed by Q. Zhou and T.-L. Ho \cite{Zhou} we can locate
the quantum critical point $\mu_c$ and then plot the ``scaled
density" $A(\mu, T)$ versus $(\mu-\mu_c)/T^{\frac{1}{\nu z}}$, where
\bea A(\mu, T)\equiv T^{-\frac{d}{z}-1+\frac{1}{z\nu}}(n(\mu,
T)-n_r). \label{density}\eea The scaled density curves for all
temperatures will collapse into a single curve. Then we can utilize
this curve of scaling function to test the critical scaling law
based on the expected critical exponents.

Within above scheme we consider the observation of the quantum
fluctuation induced first-order phase transition in $p$-orbital
bosonic system. One leading candidate to observe this phenomenon is
$^{87}$Rb atoms in a bipartite optical square lattice \cite{Wirth}.
The optical potential can be constructed by crossing two laser beams
with wavelength $\lambda=1,064$nm and $1/e^2$ radius $w_0=100 r$m.
The optical potential reads
$-\frac{V_0}{4}e^{-\frac{2z^2}{w_0^2}}|\eta[(\hat z\cos\alpha+\hat
y\sin\alpha)e^{ikx}+\epsilon\hat ze^{-ikx}]+e^{i\theta}\hat
z(e^{iky}+\epsilon e^{-iky})|^2$. $\epsilon<1$ and $\eta<1$ describe
the imperfect reflection and transmission efficiencies,
respectively. The typical values of $\epsilon$ and $\eta$ are
$\epsilon\approx0.81$ and $\eta\approx0.95$. A BEC of $2\times10^5$
$^{87}$Rb atoms (in the $F=2$, $m_F=2$ state) is produced in the
optical trap. With $V_0$ set to $V_0/E_{\mbox{rec}}=6.2$ the
excitation of $p$-orbital band can be obtained by ramping $\theta$
from $0.38\pi$ to $0.53\pi$ within 0.2ms. The initial values of the
parameter $ r_{1}$ and $ r_{2}$ can be properly chosen by tuning
$\alpha$, which is the angle between the $z$ axis and the linear
polarization of the incident beam.

We can fine-tune $\alpha$ to zero so that the system is prepared in
the vicinity of the critical line ``b1" in Fig. \ref{fig:qfisb}. By
taking \emph{in situ} measurement the density profile of the system
can be extracted. Following Q. Zhou and T.-L. Ho's scheme
\cite{Zhou} we can draw a curve of the universal scaling function.
If the system undergos a first-order phase transition the scaled
density will be in form of $A(\mu, T)=n(\mu, T)-n_r$ near the
first-order QCP, where we have used $z=2$, $d=2$ and
$\nu=\frac{1}{4}$ in Eq. (\ref{density}). To compare this phase
transition with the second-order phase transition we also calculate
the scaled density near the second-order QCP, which belongs to the
two-dimensional XY universality class with critical exponents $z=2$
and $ \nu=1/2$. Then the scaled density is $A(\mu, T)=T^{-1}(n(\mu,
T)-n_r)$. By testing which form the measured scaled density obeys we
can determine whether the phase transition is in first or second
order.

\section{Conclusion} In summary, we have investigated the quantum phase
transitions of the $p$-orbital boson gas in a square optical lattice
using renormalization group method. We find that phase transitions
from the real BEC phases to the complex BEC phase can be induced by
quantum fluctuations from the interactions between $p_x$ and $p_y$
orbital bosons. The transition indicates the $U(1)$ symmetry
breaking of $p_x$ and $p_y$ orbital bosons. This is a phenomenon
different from the $s$-orbital case, where the quantum fluctuations
tend to destroy the ordered phase but not induce it. We find that
this effect is purely induced by the inter-species interactions. Our
renormalization group analysis also indicates that this is a
first-order phase transition. Finally, we gave an experimental
proposal to observe this phenomenon in the realistic experiment.

\section{Acknowledgements}It's a pleasure to thank Congjun Wu for
suggesting the problem to us. We would also like to thank Jiangping
Hu and Anchun Ji for useful discussions. This work is supported by
the NKBRSFC under grants Nos. 2012CB821400, 2011CB921502 and
2012CB821305 and NSFC under grants Nos. 1190024, 61227902 and
61378017.

\end{document}